\input{aipcheck}
\pdfoutput=1

\documentclass[
    ,final            % use final for the camera ready runs
  ]
  {aipproc}

\layoutstyle{6x9}

\begin{document}

\title{An analytic Pad\'e-motivated QCD coupling}


\author{H. E. Mart\'inez}{
address={Physics Department, Universidad T\'ecnica Federico 
Santa Mar\'ia, Casilla 110-V, Valpara\'iso, Chile}}

\author{G. Cveti\v{c}}{
  address={Physics Department, Universidad T\'ecnica Federico Santa Mar\'ia, Casilla 110-V, Valpara\'iso, Chile}}

\begin{abstract}
We consider a modification of the Minimal Analytic (MA) coupling of Shirkov and Solovtsov. This modified MA (mMA) coupling reflects the desired analytic properties of the space-like observables. We show that an approximation by Dirac deltas of its discontinuity function $\rho$ is equivalent to a Pad\'e (rational) approximation of the mMA coupling that keeps its analytic structure. We propose a modification to mMA that, as preliminary results indicate, could be an improvement in the evaluation of low-energy observables compared with other analytic couplings.
\end{abstract}

\maketitle
%%%%%%%%%%%%%%%%%%%%%%%%%%%%%%%%%%%%%%%%%%%%
%% MAINMATTER
%%%%%%%%%%%%%%%%%%%%%%%%%%%%%%%%%%%%%%%%%%%%

%\section{Introduction}
\section{The modified MA model and its approximations}

Space-like observables $\mathcal{D}(Q^2)$ are analytic in the entire
complex $Q^2$-plane with the exception of the negative semiaxis
starting at some threshold $-M^2_0<0$. In standard perturbative QCD 
(pQCD), the coupling $a(Q^2) \equiv \alpha_s(Q^2)/\pi$
has Landau singularities at low energies 
$Q^2 \sim \Lambda^2 > 0$. This nonanalytic behavior is
reflected in the pQCD-evaluated expressions of the aforementioned
space-like observables $\mathcal{D}(Q^2)$, contravening the 
analyticity properties of the true $\mathcal{D}(Q^2)$.
The aim of analytic QCD (anQCD) models is to provide a coupling 
$A_{1}^{(an)}(Q^2)$ with the correct analytic structure.
We will consider the following expression for the QCD running 
coupling:
\begin{equation} \label{mMA}
A_{1}^{(an)}(Q^2)= \frac{1}{\pi}
\int_{M_{0}^{2}}^{\infty}d\sigma\frac{\rho_{1}(\sigma)}{\sigma+Q^2},
\end{equation} 
where $\rho_1(\sigma)={\rm Im}[a(-\sigma-i\epsilon)]$ and $a$ is
the pQCD running coupling. In the pQCD case $M_0^2=-\Lambda^2<0$.
In the widely used Minimal Analytic (MA) model \cite{Shirkov} 
$M_0^2=0$. The MA coupling is analytic down to $Q^2=0$; however, 
the latter point is not in the analyticity domain 
because the derivatives there diverge. We consider a modification 
of MA, $A_1^{(mMA)}$ with $M_0^2>0$, as introduced in
\cite{Nesterenko:2004tg}. This coupling has the same type of
analyticity domain as the space-like observables.
For the purpose of integration, $\rho_1(\sigma)$ can be 
well approximated as a sum of positively weighted Dirac deltas
(for application of such approximations in a somewhat different
context see Ref.\cite{Peris:2006ds}).
This approximation is equivalent to the paradiagonal 
Pad\'e approximant $R_{M}^{M-1}(Q^2)$ of $A_1^{(mMA)}$ 
\cite{Cvetic:2009mq}
\begin{equation} \label{narrow}
\frac{\rho_1(\sigma)}{\pi} \ \Theta(\sigma-M_0^2) \approx
\sum_{n=1}^{M}F_{n}^{2}\delta(\sigma-M_{n}^{2})
\ \Rightarrow \
A_{1}^{(mMA)}(Q^2)\approx
\sum_{n=1}^{M}\frac{F_{n}^{2}}{M_{n}^{2}+Q^2}.
\end{equation} 
We can use these approximants to evaluate the
coupling and avoid the more time-consuming calculation of the
dispersive integral in Eq. (\ref{mMA}). Further, mMA coupling is a
Stieltjes function, so all the poles $Q^2_{\rm pole}$
of its paradiagonal Pad\'e approximants lie on the negative 
real axis ($Q^2_{\rm pole} < -M_0^2$)
keeping the desired analytic structure.
\begin{figure}[htb]
\begin{minipage}[b]{.49\linewidth}
%\centering\epsfig{file=02a.jpg,width=\linewidth}
\centering\includegraphics[width=1\textwidth]{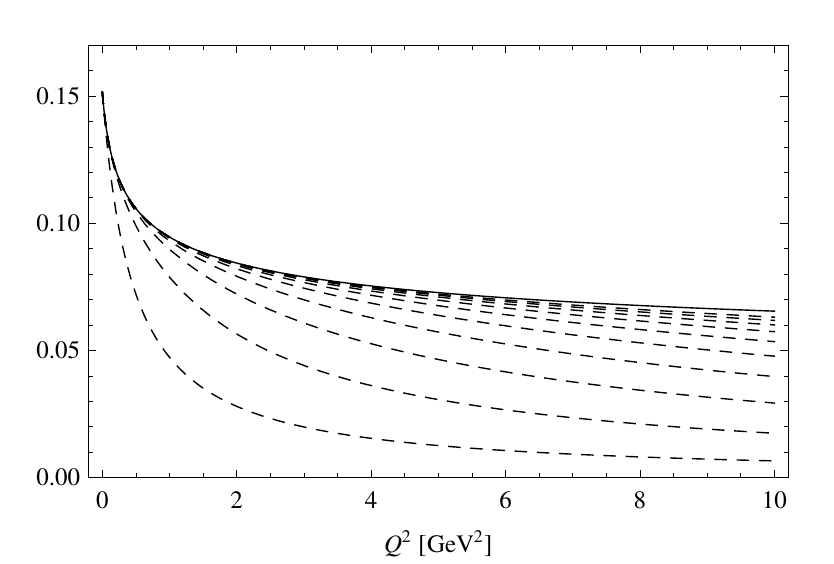}
\end{minipage}
\begin{minipage}[b]{.49\linewidth}
\centering\includegraphics[width=1\textwidth]{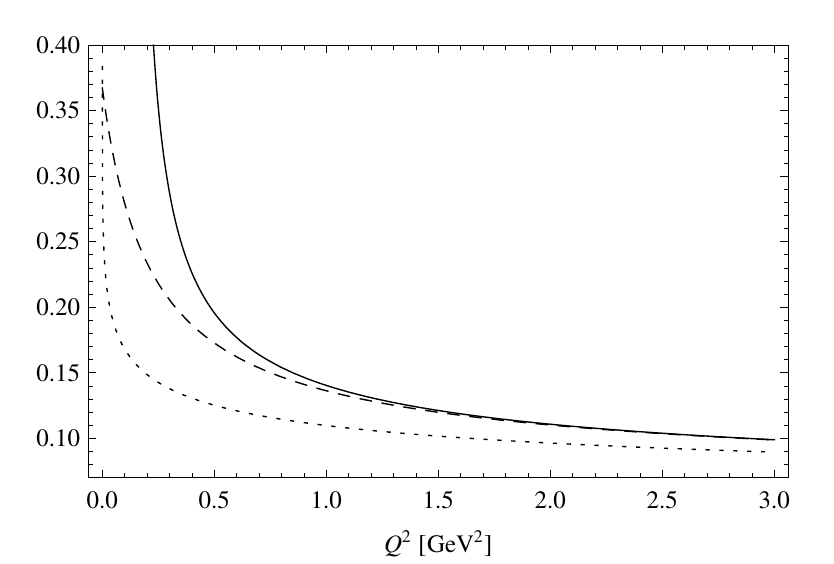}
%\centering\epsfig{file=02b.jpg,width=\linewidth}
\end{minipage}
\vspace{-0.2cm}
\caption{ Left: the full line is the mMA coupling, the dashed lines
are the Pad\'e approximants $R^{M-1}_{M}$: $R^{0}_{1}$, $R^{1}_{2}$,
..., $R^{9}_{10}$, with $M_0=2M_{\pi}$; right: pQCD coupling full 
line, MA dotted, dmMA dashed.}
\label{figure01}
\end{figure}

\section{The delta-modified MA model}

Motivated by the success of the approximation of $\rho_1$ by a
sum of Dirac deltas, and with the aim of better reproducing data,
an additional Dirac delta in Eq.~(\ref{narrow}) can be introduced
to parametrize the unknown physics in the domain of low
positive $\sigma$: 
$\Delta \rho_1(\sigma) = \pi F_{-1}^2 \delta(\sigma-M_{-1}^2)$. 
Here, $F_{-1}^2$ and $M_{-1}^2$ are new positive parameters
($0< M_{-1}^2 < M_0^2$) of this ``delta-modified'' 
MA model (dmMA). This leads to the coupling
\begin{equation} \label{dmMA}
A_1^{(dmMA)}(Q^2)=\frac{F_{-1}^2}{Q^2+M_{-1}^2}+
\frac{1}{\pi}
\int_{M_0^2}^{\infty}d \sigma \frac{\rho_1(\sigma)}{\sigma+Q^2}.
\end{equation} 
The parameters $F_{-1}^2$ and $M_{-1}^2$ can, for example, be fixed
by requiring that 
$|A_1^{(dmMA)}(Q^2)-a(Q^2)|\sim (\Lambda^2/Q^2)^3$
for $Q^2\rightarrow\infty$, 
i.e., dmMA being much closer to pQCD 
in the asymptotic region than the MA and mMA couplings ($\Rightarrow$
the value of $\Lambda^2$ practically the same as in pQCD) \cite{CCEM}. 
The remaining parameter, $M_0^2$, can be fixed by requiring that 
this coupling reproduce the experimental value of the massless 
strangeless semihadronic  $\tau$-decay ratio 
$r_{\tau}=0.202\pm 0.004$, the low-energy observable that cannot 
be reproduced correctly using MA or mMA. 

{\noindent
Work supported by Fondecyt grant 1095196 (G.C) and a PIIC-USM grant (H.M.).}
%%%%%%%%%%%%%%%%%%%%%%%%%%%%%%%%%%%%%%%%%%%%%%%%
%% BACKMATTER
%%%%%%%%%%%%%%%%%%%%%%%%%%%%%%%%%%%%%%%%%%%%%%%%

%\begin{theacknowledgments}
%This work was supported by Fondecyt grant No.~1095196 (G.C) and a PIIC-USM grant (H.M.).
%\end{theacknowledgments}



\bibliographystyle{aipproc}   % if natbib is available

\bibliography{sample}

\IfFileExists{\jobname.bbl}{}
 {\typeout{}
  \typeout{******************************************}
  \typeout{** Please run "bibtex \jobname" to optain}
  \typeout{** the bibliography and then re-run LaTeX}
  \typeout{** twice to fix the references!}
  \typeout{******************************************}
  \typeout{}
 }

\end{document}